\documentclass[prl,aps,twocolumn,superscriptaddress,showpacs]{revtex4}
\usepackage{amsmath,amssymb,graphicx}
\usepackage{pxfonts}
\usepackage{graphicx}
\usepackage{color}
\usepackage{float}
\usepackage{adjustbox}
\usepackage{graphicx}
\begin{document}

\title{Transport of Indirect Excitons in High Magnetic Fields}

\author{Y. Y.~Kuznetsova} \author{C. J.~Dorow} \author{E. V.~Calman} \author{L. V.~Butov}
\affiliation{Department of Physics, University of California at San Diego, La Jolla, California 92093-0319, USA}

\author{J.~Wilkes}
\affiliation{School of Physics and Astronomy, Cardiff University, Cardiff CF24 3AA, United Kingdom}

\author{K. L.~Campman} \author{A. C.~Gossard}
\affiliation{Materials Department, University of California at Santa Barbara, Santa Barbara, CA 93106-5050, USA}

\begin{abstract}
We present spatially- and spectrally-resolved photoluminescence measurements of indirect excitons in high magnetic fields. Long indirect exciton lifetimes give the opportunity to measure magnetoexciton transport by optical imaging. Indirect excitons formed from electrons and holes at zeroth Landau levels ($0_{\rm e} - 0_{\rm h}$ indirect magnetoexcitons) travel over large distances and form a ring emission pattern around the excitation spot. In contrast, the spatial profiles of $1_{\rm e} - 1_{\rm h}$ and $2_{\rm e} - 2_{\rm h}$ indirect magnetoexciton emission closely follow the laser excitation profile. The $0_{\rm e} - 0_{\rm h}$ indirect magnetoexciton transport distance reduces with increasing magnetic field. These effects are explained in terms of magnetoexciton energy relaxation and effective mass enhancement.
\end{abstract}


\date{\today}

\maketitle

Studies of cold fermions (electrons) in high magnetic fields have led to exciting findings, including the integer and fractional quantum Hall effects~\cite{QH}. Excitons are composite neutral bosons. The high magnetic field regime for excitons is realized when the cyclotron splitting becomes comparable to the exciton binding energy. The fulfillment of this condition for atoms requires magnetic fields $B \sim 10^6$~Tesla, and studies of cold atoms therefore use synthetic magnetic fields in rotating systems \cite{Madison00, Abo-Shaeer01, Schweikhard04} and optically synthesized magnetic fields \cite{Lin09}. Due to the small exciton mass and binding energy, the high magnetic field regime for excitons is realized with magnetic fields of a few Tesla, achievable in lab.

There are exciting theoretical predictions for cold two-dimensional (2D) neutral exciton and electron-hole (e--h) systems in high magnetic fields. Predicted collective states include a paired Laughlin liquid \cite{Yoshioka90}, an excitonic charge-density-wave state \cite{Chen91}, and a condensate of magnetoexcitons \cite{Kuramoto78, Lerner81}. Predicted transport phenomena include the exciton Hall effect \cite{Dzyubenko84, Paquet85}, superfluidity \cite{Paquet85, Imamoglu96}, and localization \cite{Dzyubenko95a}. Excitons also play an important role in the description of a many-body state in bilayer electron systems in high magnetic fields \cite{Eisenstein04}. 2D neutral exciton and e--h systems in high magnetic fields were studied experimentally in single quantum wells (QWs), and excitons and deexcitons were observed in dense e--h magnetoplasmas \cite{Butov91, Butov92}. However, low exciton temperatures were not achieved in the studied single QWs due to short exciton lifetimes.

\begin{figure}[!hb]
\includegraphics[width=\linewidth]{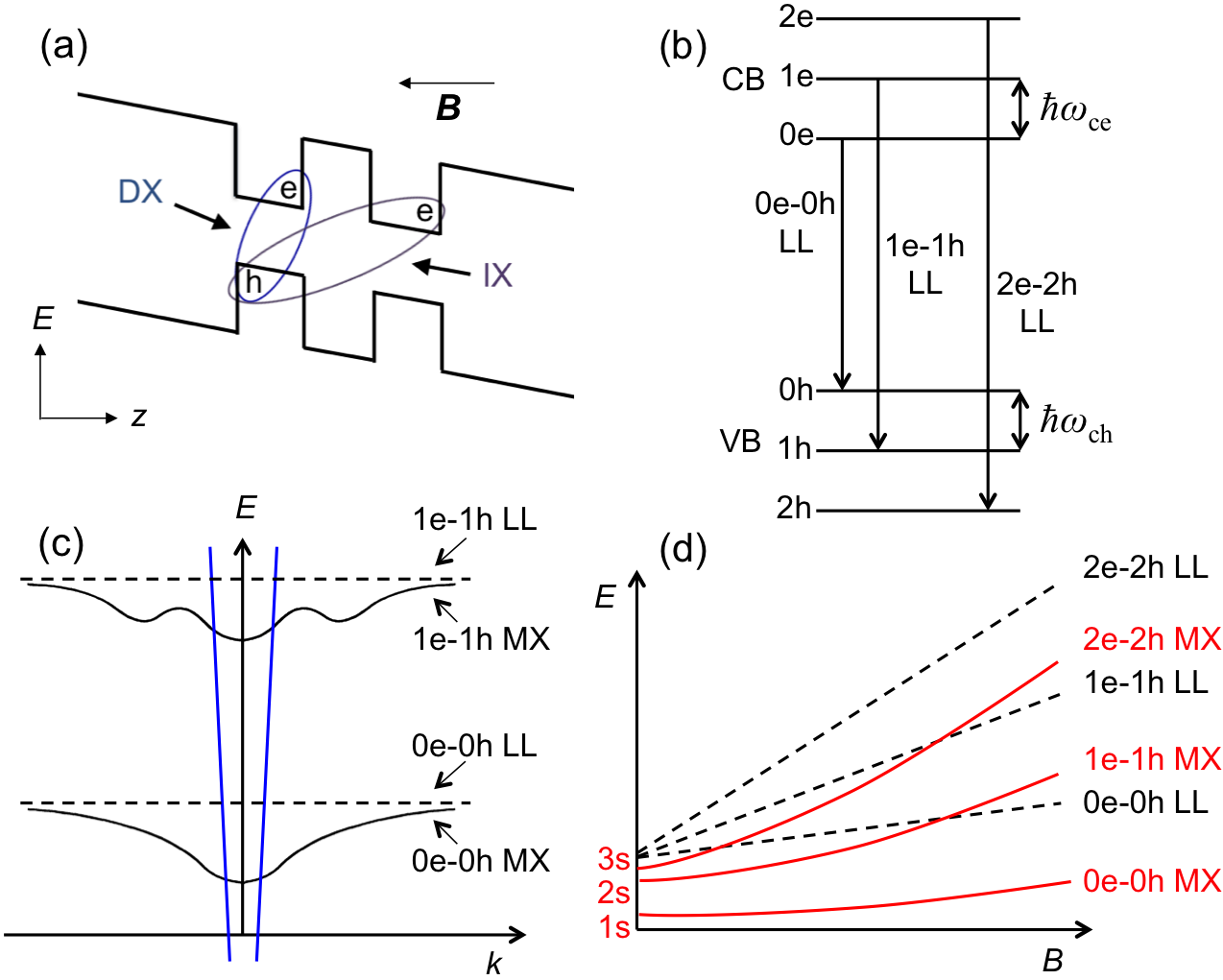}
\caption{(a) CQW band diagram. (b) Landau levels (LLs) for electron (e) in conduction band and hole (h) in valence band. Arrows show allowed optical transitions. (c) Dispersion of magnetoexciton (MX) formed from e and h at zeroth LLs, $0_{\rm e} - 0_{\rm h}$, and first LLs, $1_{\rm e} - 1_{\rm h}$ (black solid lines). Dashed lines show the sum of the e and h LL energies $1/2\hbar \omega_{\rm c}$ and $3/2\hbar \omega_c$, respectively. $\omega_{\rm c} = \omega_{\rm ce} + \omega_{\rm ch}$ is the sum of the electron and hole cyclotron energies. Blue lines show photon dispersion. (d) $k=0$ exciton energy vs magnetic field $B$ (red solid lines). Dashed black lines show the sum of the e and h LL energies $(N+1/2)\hbar \omega_{\rm c}$.}
\end{figure}

\begin{figure*}[t!]
\includegraphics[width=\textwidth]{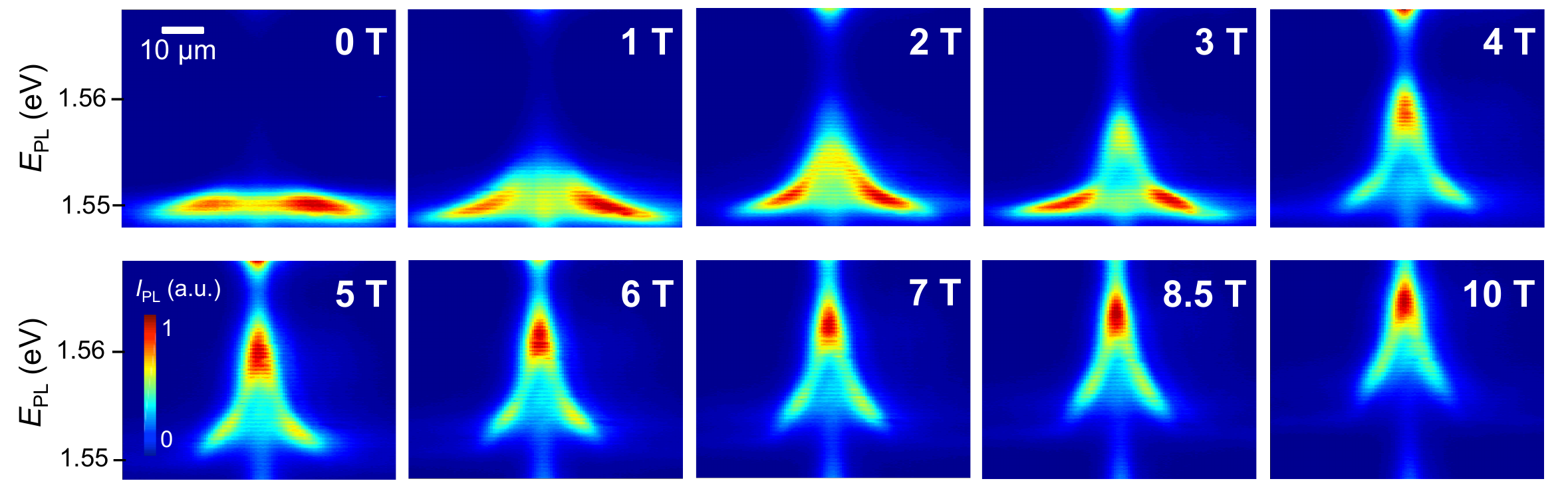}
\caption{$x-$energy IMX emission pattern for $B$ = 0 T to 10 T. Excitation power $P_{\rm ex} = 260$~$\mu$W.}
\end{figure*}

An indirect exciton (IX) in a coupled QW structure (CQW) is composed of an electron and a hole in separate QWs~\cite{Lozovik76, Fukuzawa90} (Fig.~1a). Lifetimes of IXs are orders of magnitude longer than lifetimes of regular direct excitons and long enough for the IXs to cool below the temperature of quantum degeneracy $T_0 = 2 \pi \hbar^2 n/(gk_{\rm B}M)$~\cite{Butov01} (for a GaAs CQW with the exciton spin degeneracy $g = 4$ and mass $M = 0.22m_0$, $T_0 \sim 3$~K for the exciton density per spin state $n/g = 10^{10}$~cm$^{-2}$). Furthermore, due to their long lifetimes, IXs can travel over large distances before recombination, allowing the study of exciton transport by optical imaging. Finally, the density of photoexcited e--h systems can be controlled by the laser excitation, which allows the realization of virtually any Landau level filling factor, ranging from fractional $\nu < 1$ to high $\nu$, even at fixed magnetic field. The opportunity to implement low temperatures, the high magnetic field regime, long transport distances, and controllable densities make IXs a model system for studying cold bosons in high magnetic fields.

Earlier studies of IXs in magnetic fields addressed IX energies \cite{Butov95, Dzyubenko96, Butov99, Butov01a, Kowalik-Seidl11, Schinner13}, dispersion relations \cite{Lozovik97, Butov01b, Lozovik02, Wilkes16}, and spin states \cite{Gorbunov13, High13}. Here, we present spatially- and spectrally-resolved measurements as well as theoretical simulations of IX emission, which probe IX transport and relaxation in high magnetic fields.

The theory of Mott excitons in high magnetic fields, magnetoexcitons (MXs), was developed in Refs.~\cite{Elliott, Hasegawa, Gor'kov68, Lerner80, Kallin84}. 2D MXs are shown schematically in Fig.~1. Optically active MXs are formed from electrons and holes at Landau levels (LLs) with $N_{\rm e} = N_{\rm h}$ (Fig.~1b,c). The MX dispersion is determined by the coupling between the MX center-of-mass motion and internal structure: MX is composed of an electron and a hole forced to travel with the same velocity, producing on each other a Coulomb force that is canceled by the Lorentz force. At small momenta $k$, the magnetoexciton dispersion is parabolic and can be described by an effective MX mass. At high $k$, the MX energy tends toward the sum of the electron and hole LL  energies (Fig.~1c). The MX mass and MX binding energy increase with $B$~\cite{Gor'kov68, Lerner80, Kallin84, Paquet85}. With reducing $B$, $N_{\rm e} - N_{\rm h}$ MX states transform to $(N+1)$s exciton states (Fig.~1d). These properties are characteristic of both direct MXs (DMXs) and indirect MXs (IMXs).
Due to the separation ($d$) between the e and h layers, IMX energies are lower by $\sim edF_z$ and grow faster with $B$ \cite{Butov95, Dzyubenko96, Butov99, Butov01a, Kowalik-Seidl11, Schinner13, Lozovik97} ($F_z$ is the electric field in the $z$ direction), IMX binding energies are smaller \cite{Lozovik97, Dzyubenko96, Butov01b, Lozovik02, Wilkes16}, and IMX effective masses grow faster with $B$ \cite{Lozovik97, Butov01b, Lozovik02, Wilkes16}.

Free 2D MXs can recombine radiatively when their momentum $k$ is inside the intersection between the dispersion surface $E_{\rm MX}(k)$ and the photon cone $E = \hbar k c / \sqrt{\varepsilon}$, called the radiative zone \cite{Feldman87, Hanamura88, Andreani91} (Fig.~1c). In GaAs QW structures, the radiative zone corresponds to $k \lesssim k_0 \approx E_{\rm g} \sqrt{\varepsilon} / (\hbar c) \approx 2.7 \cdot 10^5$~cm$^{-1}$ ($\varepsilon$ is the dielectric constant, $E_{\rm g}$ the semiconductor gap). In GaAs QW structures, excitons may have four spin projections on the $z$ direction $J_z = \pm 2, \pm 1$; the $J_z = \pm 1$ states are optically active \cite{Maialle93}. Free MXs with $k \lesssim k_0$, $N_{\rm e} = N_{\rm h}$, and $J_z = \pm 1$ recombine radiatively directly contributing to MX emission. Free MXs with $k > k_0$, $N_{\rm e} \ne N_{\rm h}$, or $J_z = \pm 2$ are dark.

Experiments were performed on a $n^+ - i - n^+$ GaAs CQW. The $i$-region consists of a single pair of 8-nm GaAs QWs separated by a 4-nm Al$_{0.33}$Ga$_{0.67}$As barrier, surrounded by 200-nm Al$_{0.33}$Ga$_{0.67}$As layers. The $n^+$ layers are Si-doped GaAs with Si concentration $5 \cdot 10^{17}$~cm$^{-3}$. The indirect regime, characterized by IX being the lowest energy state, was implemented by applying voltage $V = - 1.2$~V between the $n^+$ layers. The 633~nm laser excitation was focused to $\sim 6$~$\mu$m spot. The $x-$energy images were measured with a liquid-nitrogen-cooled charge-coupled device (CCD) placed after a spectrometer with resolution 0.18~meV. Spatial resolution was $\approx 2.5$~$\mu$m. The measurements were performed in an optical dilution refrigerator at temperature $T_{\rm bath} = 40$~mK and magnetic fields $B = 0 - 10$~T perpendicular to the CQW plane.

Figure~2 shows the evolution of measured $x-$energy emission patterns with increasing $B$. Horizontal cross sections of the $x-$energy emission pattern reveal the spatial profiles at different energies (Fig.~3a,b), while vertical cross sections present spectra at different distances $x$ from the excitation spot center (Fig.~3c,d). More spatial profiles and spectra of the IMX emission at different $B$ and $P_{\rm ex}$ are presented in \cite{suppl_mat}. Spatial profiles of the amplitudes of IMX emission lines are shown in Fig.~4a,b.

\begin{figure}[t!]
\includegraphics[width=\linewidth]{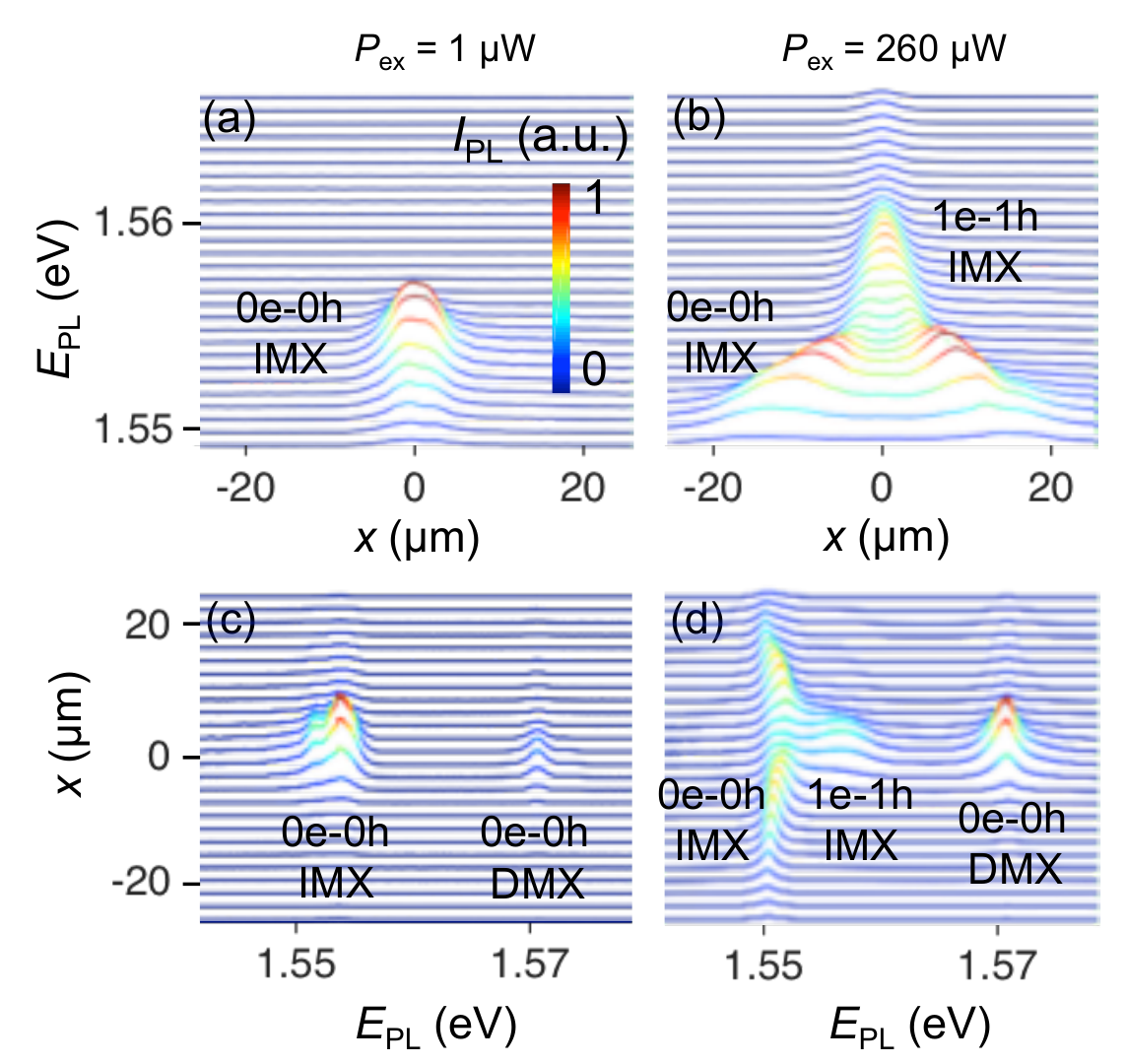}
\caption[width=\linewidth]{(a,b) Spatial profiles of the IMX emission at different energies at $P_{\rm ex} = 1$ (a) and 260~$\mu$W (b). (c,d) MX spectra at different distances $x$ from the excitation spot center at $P_{\rm ex} = 1$ (c) and 260~$\mu$W (d). IMX and DMX lines correspond to indirect and direct MX emission, respectively. The low-energy bulk emission was subtracted from the spectra \cite{suppl_mat}. The laser excitation profile is shown in Fig.~4a. $B$ = 3 T.}
\end{figure}

\begin{figure}[t!]
\includegraphics[width=\linewidth]{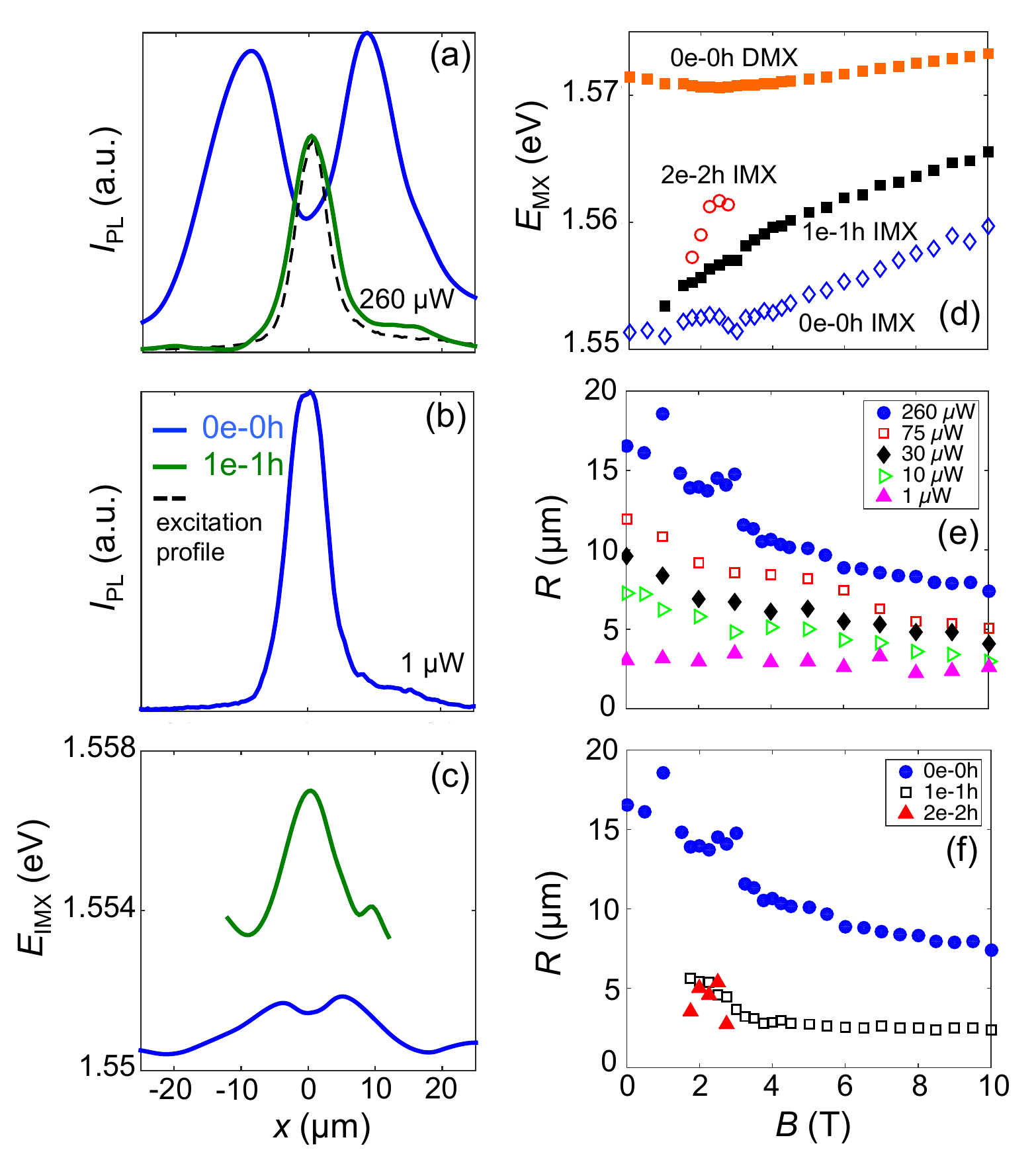}
\caption{(a,b) Amplitude of $0_{\rm e} - 0_{\rm h}$ and $1_{\rm e} - 1_{\rm h}$ IMX emission lines at $P_{\rm ex} = 260$ (a) and 1~$\mu$W (b), $B = 3$~T. (c) $0_{\rm e} - 0_{\rm h}$ and $1_{\rm e} - 1_{\rm h}$ IMX energies at $P_{\rm ex} = 260$~$\mu$W, $B$ = 3~T. (d) IMX energies vs $B$ at $P_{\rm ex} = 260$~$\mu$W and $x$ = 0. (e) $0_{\rm e} - 0_{\rm h}$ IMX emission radius $R$ (half-width-half-maximum) vs $B$ for different $P_{\rm ex}$. (f) $0_{\rm e} - 0_{\rm h}$, $1_{\rm e} - 1_{\rm h}$, and $2_{\rm e} - 2_{\rm h}$ IMX emission radius vs $B$ for $P_{\rm ex} = 260$~$\mu$W. Laser excitation profile shown in (a).}
\end{figure}

The magnetic field dependence (Fig.~4d) identifies the $0_{\rm e} - 0_{\rm h}$, $1_{\rm e} - 1_{\rm h}$, and $2_{\rm e} - 2_{\rm h}$ IMX emission lines. The DMX emission is also observed at high energies (Figs.~3, 4d).

At low $P_{\rm ex}$ and, in turn, low IMX densities, the $0_{\rm e} - 0_{\rm h}$ IMX emission essentially follows the laser excitation profile (Figs.~3a,c and 4b). This indicates that at low densities, $0_{\rm e} - 0_{\rm h}$ IMXs are localized in the in-plane disorder potential and practically do not travel beyond the excitation spot. However, at high densities, transport of $0_{\rm e} - 0_{\rm h}$ IMXs is observed as the $0_{\rm e} - 0_{\rm h}$ IMX emission extends well beyond the excitation spot (Figs.~2, 3b,d, and 4a). Furthermore, the $0_{\rm e} - 0_{\rm h}$ IMX emission shows a ring structure around the excitation spot (Figs.~2, 3b,d, and 4a). This structure is similar to the inner ring in the IX emission pattern at $B = 0$~\cite{Ivanov06, Hammack09, Kuznetsova12}. The enhancement of $0_{\rm e} - 0_{\rm h}$ IMX emission intensity with increasing distance from the center originates from IMX transport and energy relaxation as follows. IMXs cool toward the lattice temperature when they travel away from the laser excitation spot, thus forming a ring of cold IMXs. The cooling increases the occupation of the low-energy optically active IMX states (Fig.~1c), producing the $0_{\rm e} - 0_{\rm h}$ IMX emission ring. The ring extension $R$ characterizing the $0_{\rm e} - 0_{\rm h}$ IMX transport distance is presented in Fig.~4e,f. The IMX transport distance increases with density (Fig.~4e,f). This effect is explained by the theory presented below in terms of the screening of the structure in-plane disorder by the repulsively interacting IMXs.

In contrast to the $0_{\rm e} - 0_{\rm h}$ IMX emission, the spatial profile of the $1_{\rm e} - 1_{\rm h}$ IMX emission closely follows the laser excitation profile (Fig.~4a). The data show that the high-energy $1_{\rm e} - 1_{\rm h}$ IMX states are occupied in the excitation spot region (where the IMX temperature and density is maximum). Long transport is not observed here because the $1_{\rm e} - 1_{\rm h}$ IMXs effectively relax in energy and transform to $0_{\rm e} - 0_{\rm h}$ IMXs beyond the laser excitation spot where the IMX temperature drops down. $1_{\rm e} - 1_{\rm h}$ IMX transport distance within this relaxation time $\lesssim 3$~$\mu$m (Fig.~4a,f).

Additionally, the $0_{\rm e} - 0_{\rm h}$ and $1_{\rm e} - 1_{\rm h}$ IMX energies are observed to reduce with $x$ (Fig.~4c). This energy reduction follows the IMX density reduction away from the excitation spot. The density reduction can lower IX energy due to interactions and localization in the disorder potential. IMXs have a built-in dipole moment $\sim ed$ and interact repulsively. The repulsive IMX interaction causes the reduction of the IMX energy with reducing density. In contrast, direct MXs in single QWs are essentially noninteracting particles and their energy practically does not depend on density \cite{Lerner81, Butov92}.

Figure~4e,f also shows that the $0_{\rm e} - 0_{\rm h}$ IMX transport distance reduces with $B$. This effect is explained below in terms of the enhancement of the MX mass. The reduction of the IMX transport distance causes the IMX accumulation in the excitation spot area. The IMX accumulation contributes to the observed enhancement of both the IMX emission intensity and energy in the excitation spot area with increasing $B$ (Fig.~2).

We note also that IX emission patterns may contain the inner ring, which forms due to IX transport and thermalization \cite{Ivanov06, Hammack09, Kuznetsova12}, and external ring, which forms on the interface between the hole-rich and electron-rich regions \cite{Butov04, Rapaport04, Chen05, Haque06, Yang10}. The data presented in Fig.~S1 \cite{suppl_mat} show that the external ring and the presence of the charge-rich regions associated with it play no major role in the IMX transport and relaxation phenomena described in this work.

\begin{figure}[t!]
\includegraphics[width=\linewidth]{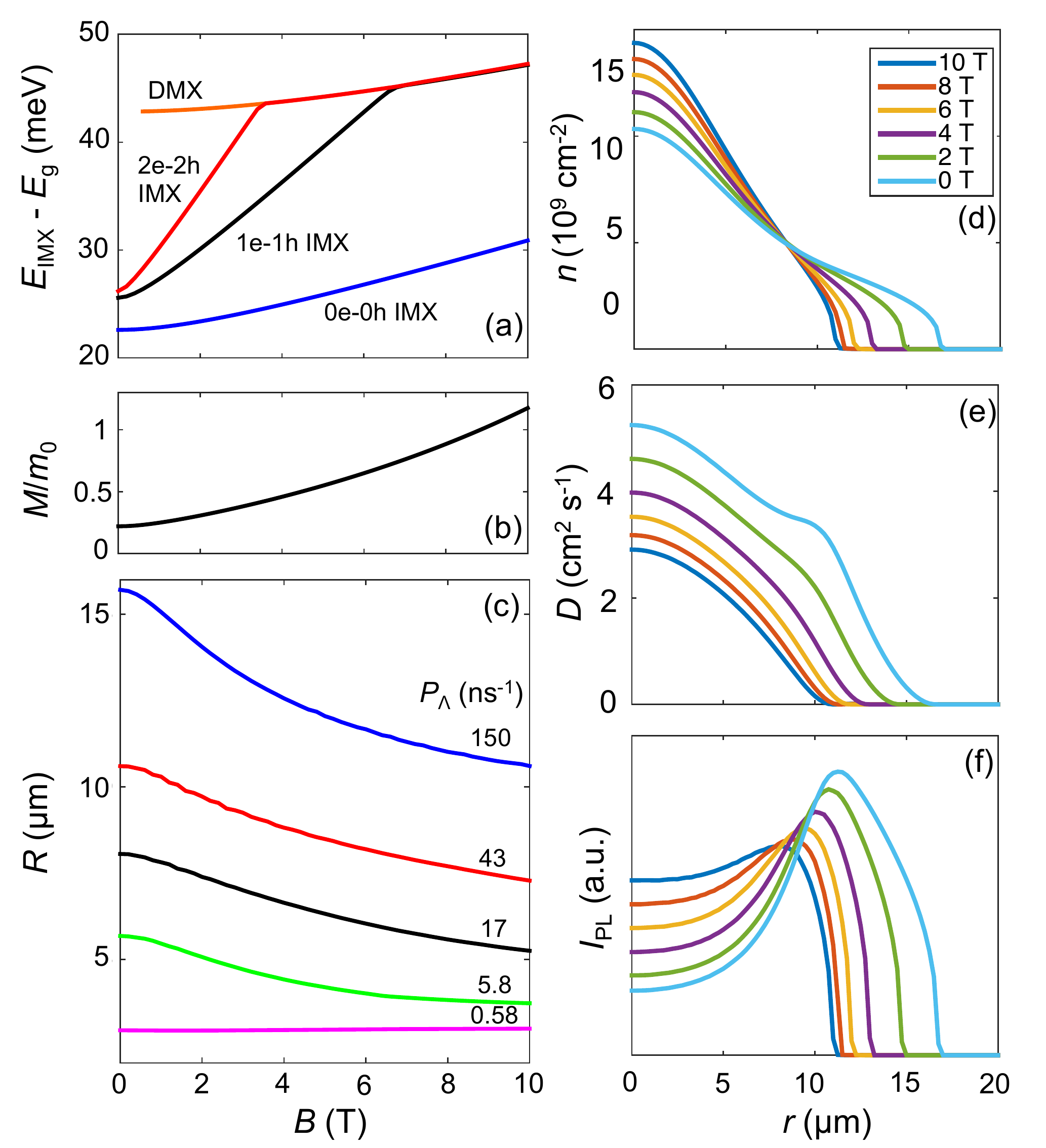}
\caption{(a) Optical transition energies of $k=0$ single MX states measured from the band gap vs $B$ calculated using a multi-sub-level approach. (b) $0_{\rm e} - 0_{\rm h}$ IMX mass renormalization due to the magnetic field calculated via perturbation theory. (c) Simulated ring radius $R$ (defined as HWHM of $I_{\rm PL}$ shown in (f)) vs $B$ for different injection rates, $P_\Lambda=2\pi\int_0^\infty\Lambda(r)r\,dr$. Spatial profiles of the (d) density, (e) diffusion coefficient and (f) emission intensity of $0_{\rm e} - 0_{\rm h}$ IMX from solving Eqns.\,(\ref{transportEqn})-(\ref{thermalizationEqn}) for different $B$ with injection rate $150\,{\rm ns^{-1}}$ and $T_{\rm bath} = 0.5$~K.}
\end{figure}

The following two-body Hamiltonian is used to describe MXs in CQWs under external bias,
\begin{equation}
\hat{H}({\bf r}_e,{\bf r}_h) = \hat{H}_e({\bf r}_e) + \hat{H}_h({\bf r}_h) - \frac{e^2}{\varepsilon|{\bf r}_e - {\bf r}_h|} + E_g. \label{Hamiltonian}
\end{equation}
Here, ${\bf r}_{e(h)}$ and $\hat{H}_{e(h)}$ are the electron (hole) coordinates and single-particle Hamiltonians, respectively. The latter are given by
\begin{equation}
\hat{H}_{e(h)}({\bf r}) = \hat{p}_{e(h)}({\bf r})\frac{1}{2}\hat{m}_{e(h)}^{-1}(z)\hat{p}_{e(h)}({\bf r}) + U_{e(h)}(z).
\end{equation}
The magnetic field $B$ contributes to the momentum operators $\hat{p}_{e(h)}({\bf r}) = -i\hbar\nabla_{\bf r} \pm (e/c) {\bf A}({\bf r})$ via the magnetic vector potential $\bf A$. The mass tensor $\hat{m}_{e(h)}(z)$ contains the electron (hole) effective masses which are step functions along $z$ due to the electron and hole confinement in the corresponding QW layers (Fig.~1a). $U_{e(h)}(z)$ contain the QW confinement and the potential due to the applied electric field. The third term in Eq.~\ref{Hamiltonian} is the e-h Coulomb interaction. After a factorization of the wave function to separate the center of mass and relative coordinates~\cite{Lozovik97}, eigen states of the Hamiltonian describing the relative motion of an exciton with $k = 0$ are found using a multi-sub-level approach~\cite{Sivalertporn12, Wilkes16}. This allows the extraction of the $B$-field dependence of the $k = 0$ IMX energy, $E_{\rm IMX}$, and radiative lifetime, $\tau_R$. Treating $k$ as a perturbation, we then use perturbation theory to second order to determine the exciton in-plane effective mass enhancement due to the magnetic field, $M(B)$. The calculated $E_{\rm IMX}$ and $M(B)$ are in agreement with the measured $E_{\rm IMX}$ (compare Fig.~5a and 4d) and $M(B)$ (compare Fig.~5b and $M(B)$ in Ref.~\cite{Butov01b, Lozovik02}).

The $B-$dependence of the ring in the IMX emission pattern is simulated by combining the microscopic description of a single IMX with a model of IMX transport and thermalization. The following set of coupled equations were solved for the $0_{\rm e} - 0_{\rm h}$ IMX density $n(r,t)$ and temperature $T(r,t)$ in the space-time ($r$,$t$) domain,
\begin{eqnarray}
\frac{\partial n}{\partial t} = \nabla \left[D\nabla n + \mu_{\rm x} n \nabla (u_0 n)\right] + \Lambda - \frac{n}{\tau}, \label{transportEqn} \\
\frac{\partial T}{\partial t} = S_{\rm pump} - S_{\rm phonon}. \label{thermalizationEqn}
\end{eqnarray}
The two terms in square brackets in the transport equation (\ref{transportEqn}) describe IMX diffusion and drift currents. The latter originates from the repulsive dipolar interactions approximated by $u_0 = 4 \pi e^2 d/\varepsilon$ within the model~\cite{Hammack09}. The diffusion coefficient $D$ and mobility $\mu_{\rm x}$ are related by a generalized Einstein relation, $\mu_{\rm x} = D(e^{T_0/T} - 1)/(k_BT_0)$. An expression for $D$ is derived using a thermionic model to account for the screening of the random QW disorder potential by dipolar excitons \cite{Ivanov06, Hammack09, Kuznetsova12}. $D$ is inversely proportional to the exciton mass $M$. The enhancement of $M$ with $B$ describes the magnetic field induced reduction in exciton transport. The last two terms on the RHS of (\ref{transportEqn}) describe creation and decay of excitons. $\Lambda(r)$ has a Gaussian profile chosen to match the excitation beam. The optical lifetime $\tau$ is the product of $\tau_R$ and a factor that accounts for the fraction of excitons that are inside the radiative zone.
The effects of IMX in higher levels are included via $\Lambda$, since they relax to the $0_{\rm e} - 0_{\rm h}$ level within the excitation region.

The thermalization equation (\ref{thermalizationEqn}) describes the balance between heating of excitons by non-resonant photoexcitation and cooling via interaction with bulk longitudinal acoustic phonons. Both rates are modified by the magnetic field due to their dependence on $M(B)$. The emission intensity is extracted from $n/\tau$. In the simulations, $T_{\rm bath} = {\rm 0.5\,K}$ was used to avoid the excessive computation times incurred by the dense grids needed to handle the strongly non-linear terms in (\ref{transportEqn})-(\ref{thermalizationEqn}) that are most prominent at low $T$. Below $T_{\rm bath}$ $\approx {\rm 1\,K}$, the results of the model are qualitatively similar with the ring radius only slowly varying with $T$. Modifying the computations for lower $T_{\rm bath}$ forms the subject of future work. Details of the transport and thermalization model including parameters and expressions for $D$, $\tau$, $S_{\rm pump}$ and $S_{\rm phonon}$ can be found in~\cite{suppl_mat}.

The simulations show the ring in the IMX emission pattern (Fig.~5f) in agreement with the experiment (Fig.~2, 3b,d, 4a). The increase of the IMX mass causes the reduction of the IMX diffusion coefficient (Fig.~5e), contributing to the reduction of the IMX transport distance with magnetic field (Fig.~5c). The measured and simulated reduction of the IMX transport distance with magnetic field are in agreement (compare Fig.~4e and 5c).

This work was supported by NSF Grant No. 1407277. J.W. was supported by the EPSRC (grant EP/L022990/1). C.J.D. was supported by the NSF Graduate Research Fellowship Program under Grant No. DGE-1144086.

\begin{figure*}[htbp]
\vspace*{-5mm}
\vspace{-\parskip}
\vspace{-\topskip}
 \hspace*{-15mm}
\includegraphics[width=1.16\textwidth]{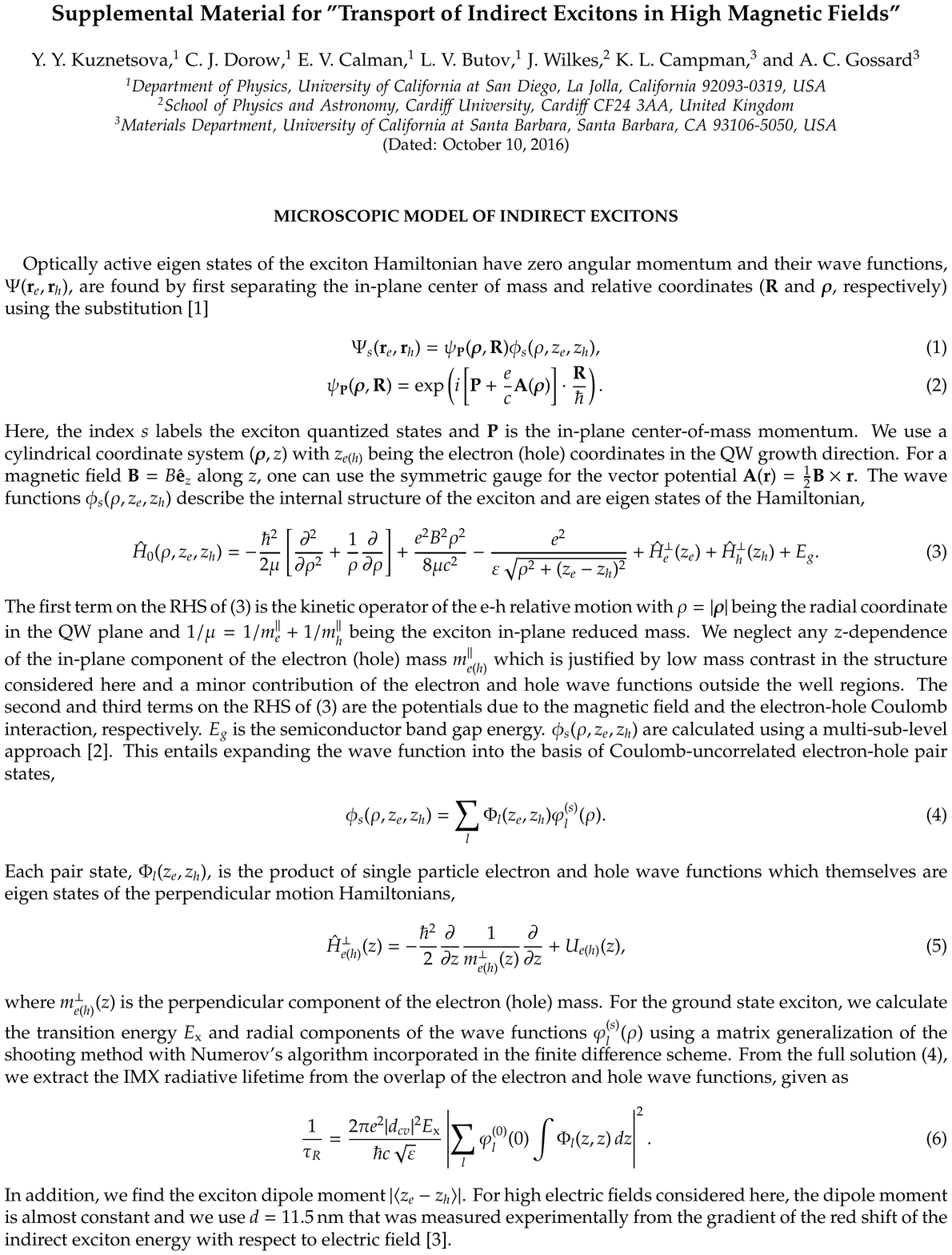}
\end{figure*}

\null
\begin{figure*}[htbp]
\vspace*{-5mm}
\vspace{-\parskip}
\vspace{-\topskip}
 \hspace*{-15mm}
\includegraphics[width=1.16\textwidth]{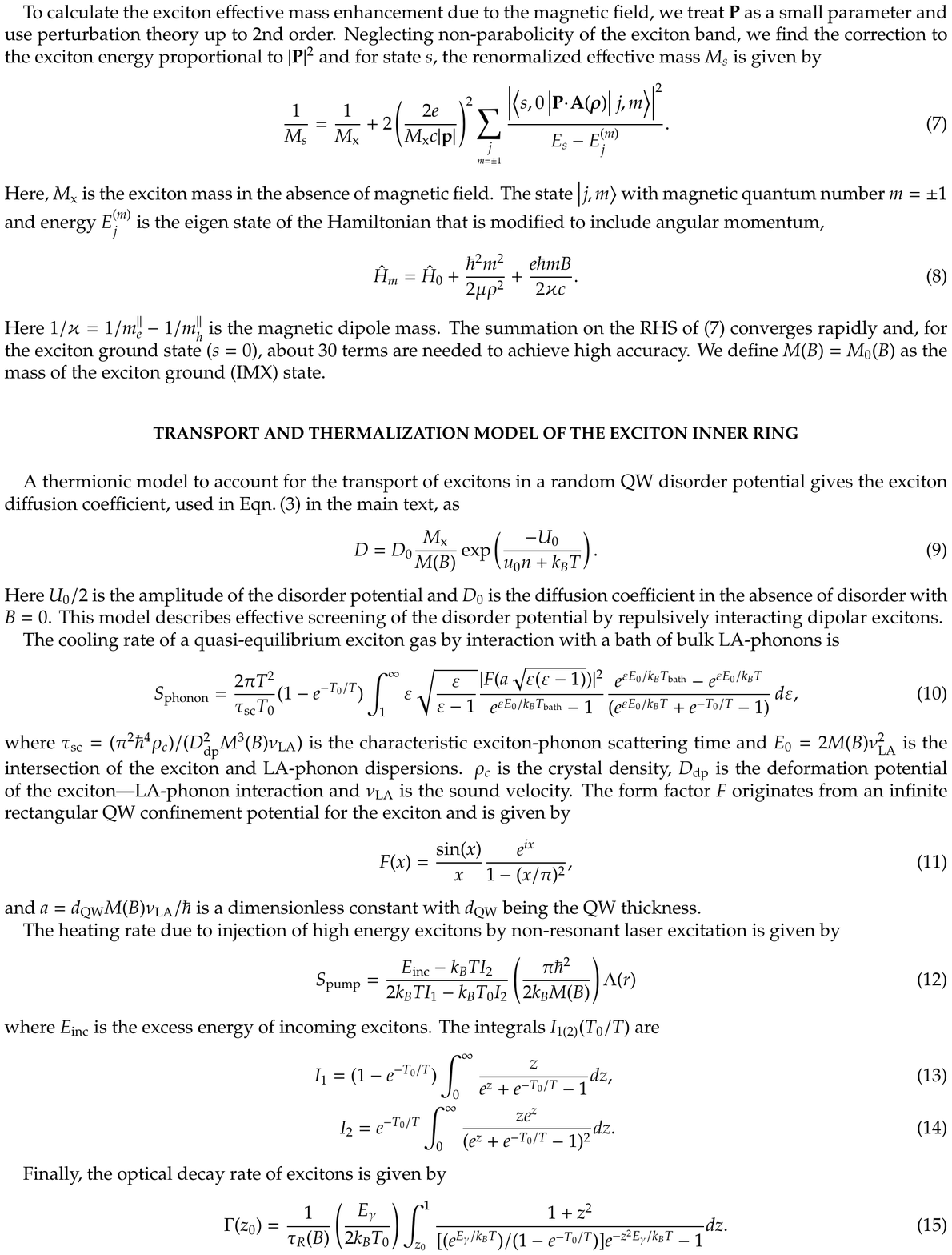}
\end{figure*}

\begin{figure*}[htbp]
\vspace*{-5mm}
\vspace{-\parskip}
\vspace{-\topskip}
 \hspace*{-15mm}
\includegraphics[width=1.16\textwidth]{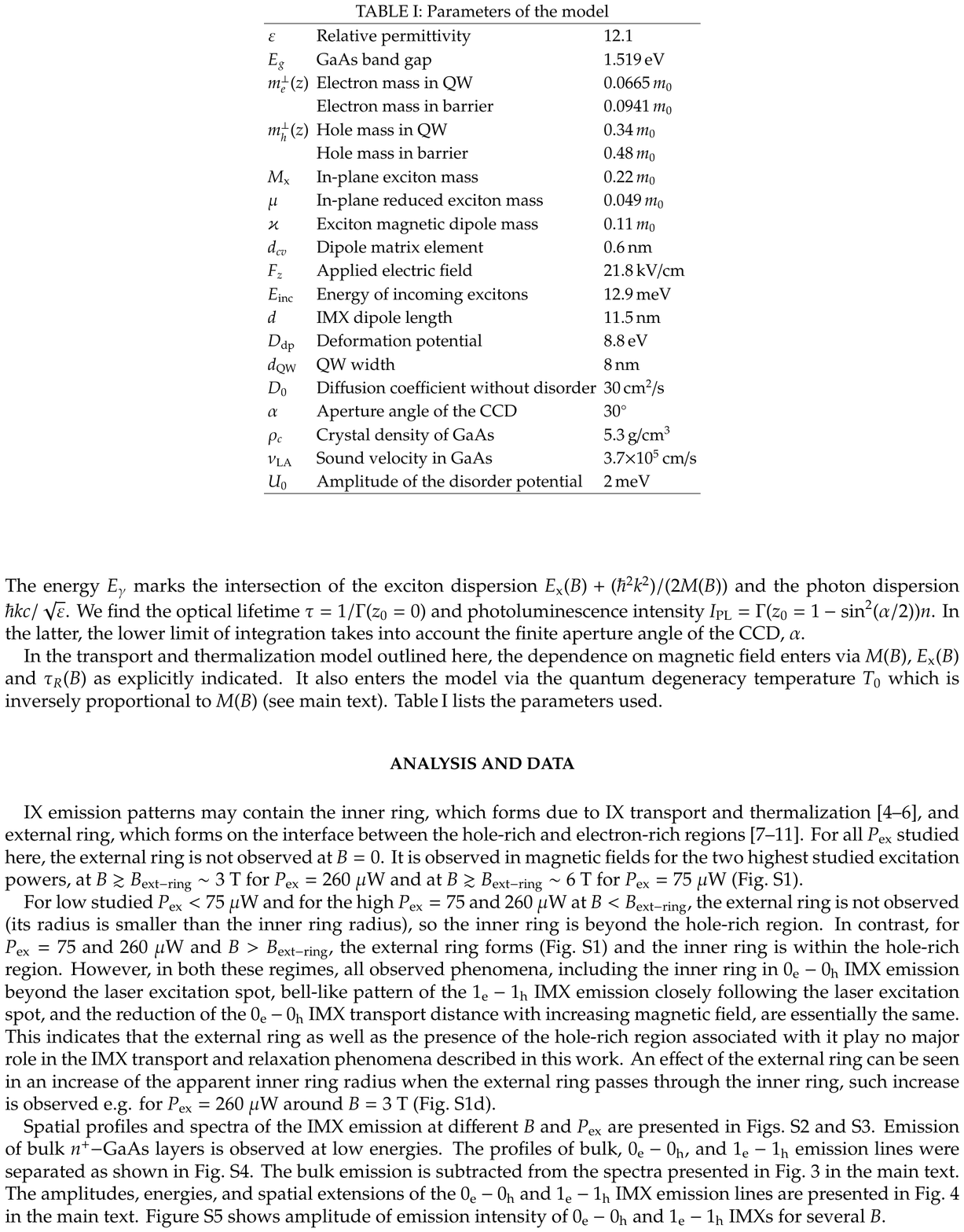}
\end{figure*}

\begin{figure*}[htbp]
\vspace*{-5mm}
\vspace{-\parskip}
\vspace{-\topskip}
 \hspace*{-15mm}
\includegraphics[width=1.16\textwidth]{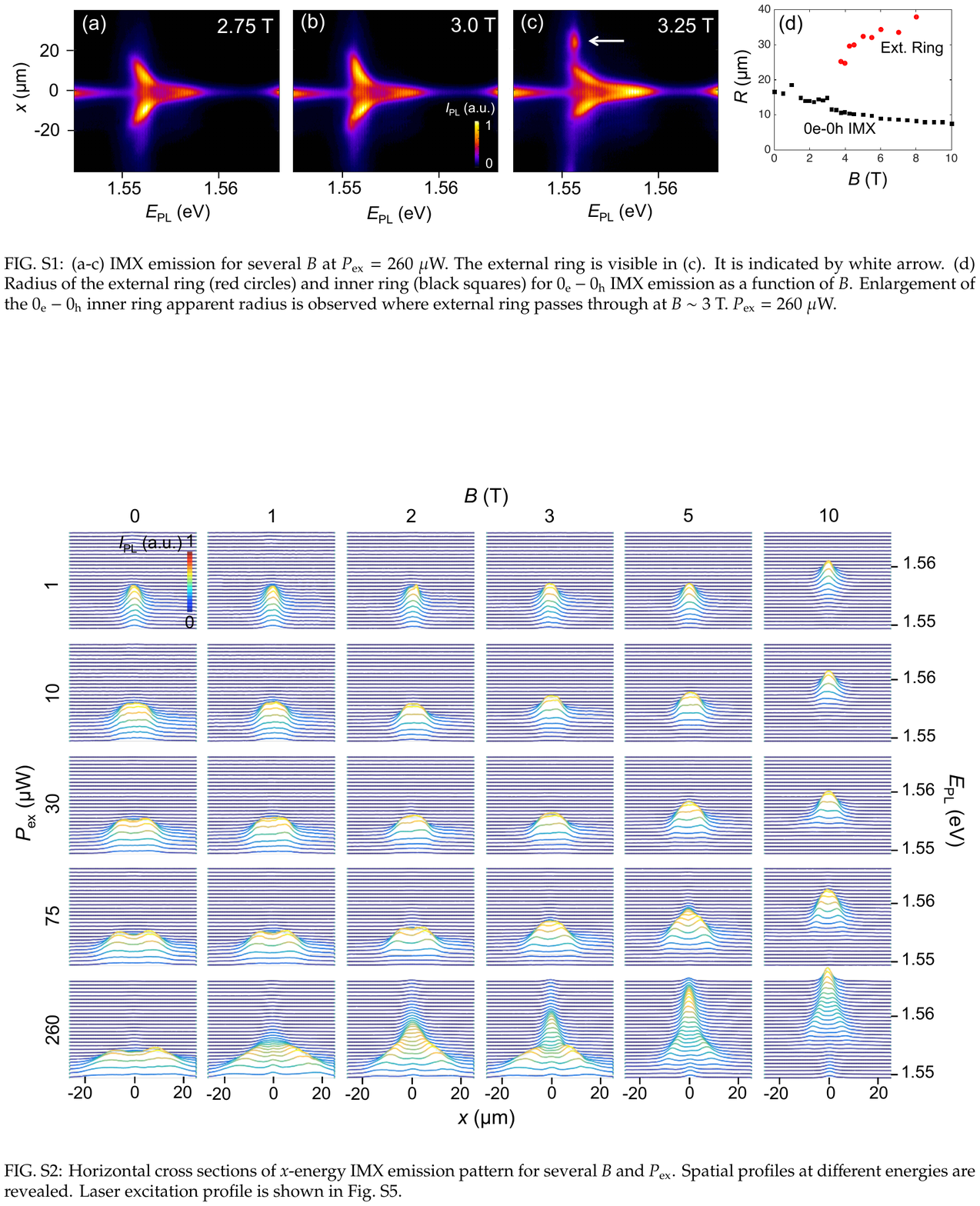}
\end{figure*}

\begin{figure*}[htbp]
\vspace*{-5mm}
\vspace{-\parskip}
\vspace{-\topskip}
 \hspace*{-15mm}
\includegraphics[width=1.16\textwidth]{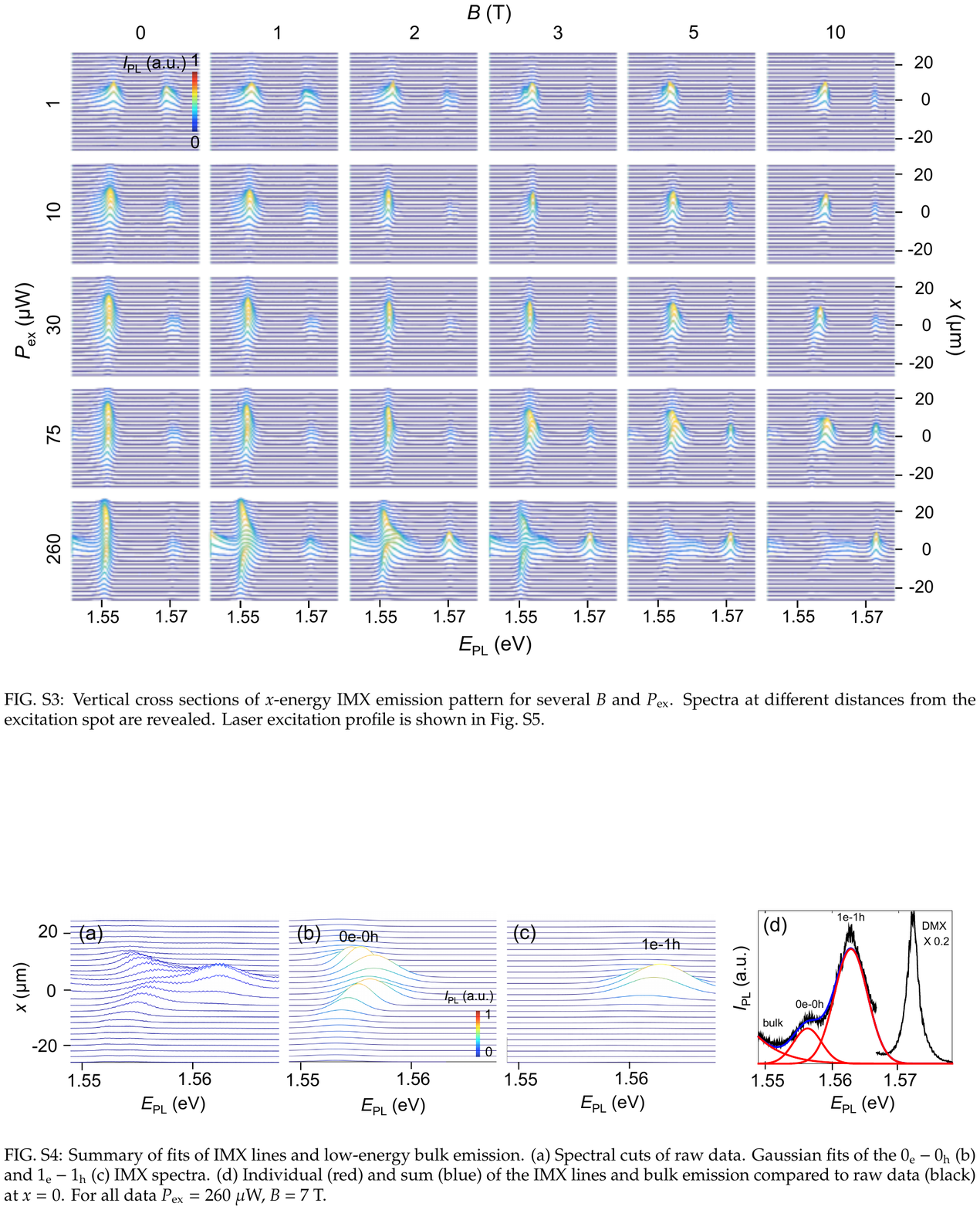}
\end{figure*}

\begin{figure*}[htbp]
\vspace*{-5mm}
\vspace{-\parskip}
\vspace{-\topskip}
 \hspace*{-15mm}
\includegraphics[width=1.16\textwidth]{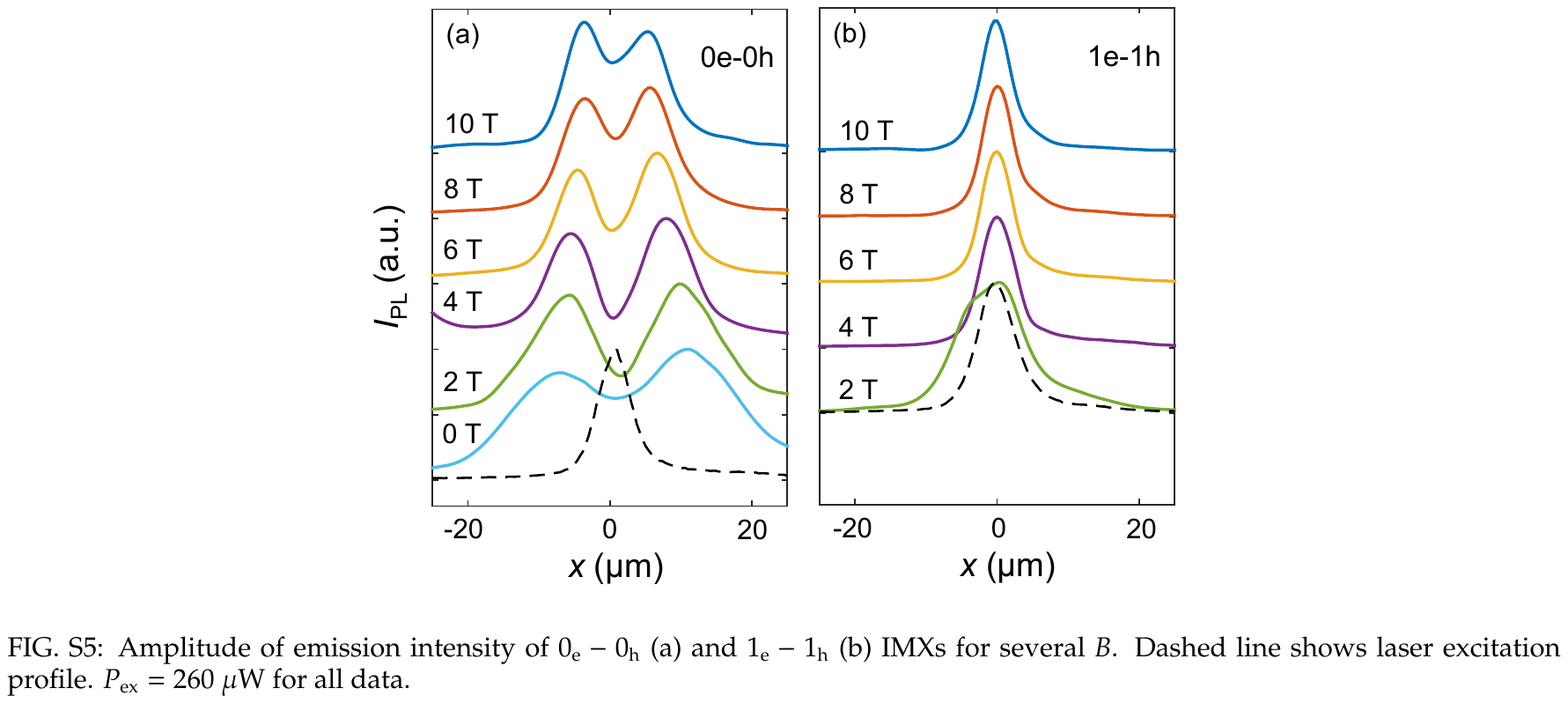}
\end{figure*}

\begin{figure*}[htbp]
\vspace*{-5mm}
\vspace{-\parskip}
\vspace{-\topskip}
 \hspace*{-15mm}
\includegraphics[width=1.16\textwidth]{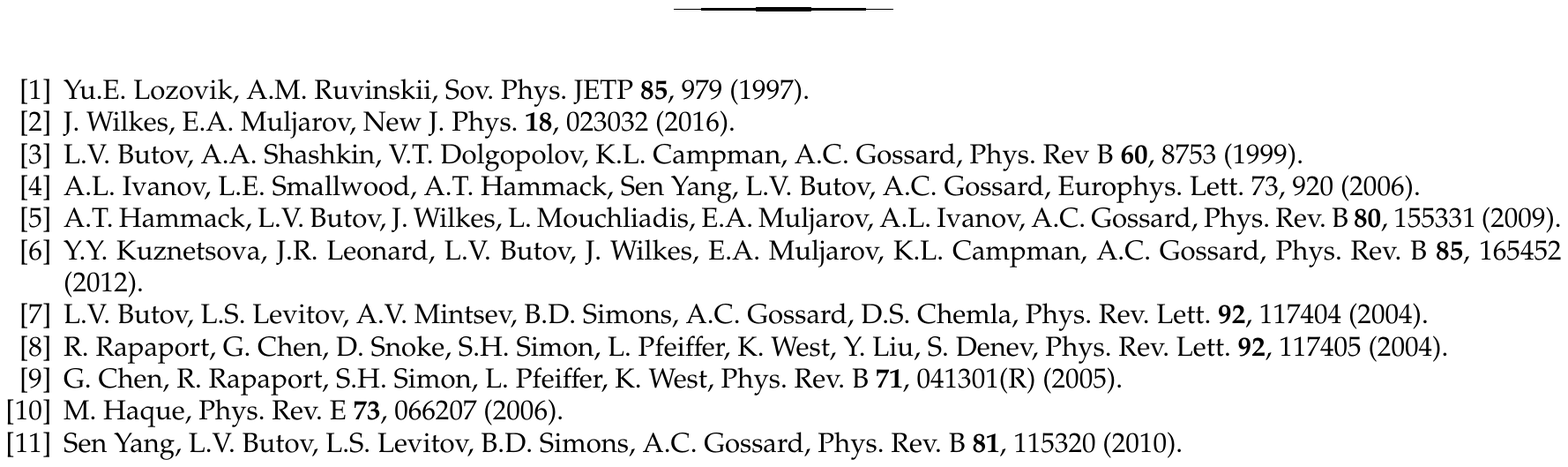}
\end{figure*}


\begin{references}

\bibitem{QH}
For reviews, see {\it Perspectives in Quantum Hall Effects}, Ed. S. Das Sarma, A. Pinczuk (Wiley, New York, 1997);
H.L. St{\"o}rmer, D.C. Tsui, A.C. Gossard,
Rev. Mod. Phys. {\bf 71}, S298
(1999).

\bibitem{Madison00}
K.W. Madison, F. Chevy, W. Wohlleben, J. Dalibard, Phys. Rev. Lett. {\bf 84}, 806 (2000).

\bibitem{Abo-Shaeer01}
J.R. Abo-Shaeer, C. Raman, J.M. Vogels, W. Ketterle, Science {\bf 292}, 476 (2001).

\bibitem{Schweikhard04}
V. Schweikhard, I. Coddington, P. Engels, V.P. Mogendorff, E.A. Cornell, Phys. Rev. Lett. {\bf 92}, 040404 (2004).

\bibitem{Lin09}
Y.-J. Lin, R.L. Compton, K. Jim{\'e}nez-Garc{\'i}a, J.V. Porto, I.B. Spielman, Nature {\bf 462}, 628 (2009).

\bibitem{Yoshioka90}
D. Yoshioka, A.H. MacDonald,
J. Phys. Soc. Jpn. {\bf 59}, 4211 (1990).

\bibitem{Chen91}
X.M. Chen, J.J. Quinn, Phys. Rev. Lett. {\bf 67}, 895 (1991).

\bibitem{Kuramoto78}
Y. Kuramoto, C. Horie,
Solid State Commun. {\bf 25}, 713
(1978).

\bibitem{Lerner81}
I.V. Lerner, Yu.E. Lozovik, Sov. Phys. JETP {\bf 53}, 763 (1981).

\bibitem{Dzyubenko84}
A.B. Dzyubenko, Yu.E. Lozovik,
Sov. Phys. Solid State {\bf 26}, 938 (1984).

\bibitem{Paquet85}
D. Paquet, T.M. Rice, K. Ueda, Phys. Rev. B {\bf 32}, 5208 (1985).

\bibitem{Imamoglu96}
A. Imamoglu, Phys. Rev. B {\bf 54}, R14285 (1996).

\bibitem{Dzyubenko95a}
P.I. Arseyev, A.B. Dzyubenko,
Phys. Rev. B {\bf 52}, R2261
(1995).

\bibitem{Eisenstein04}
J.P. Eisenstein, A.H. MacDonald,
Nature {\bf 432}, 691
(2004).

\bibitem{Butov91}
L.V. Butov, V.D. Kulakovskii, E.I. Rashba, Sov. Phys. JETP Lett. {\bf 53}, 109 (1991).

\bibitem{Butov92}
L.V. Butov, V.D. Kulakovskii, G.E.W. Bauer, A. Forchel, D. Gr{\"u}tzmacher, Phys. Rev. B {\bf 46}, 12765 (1992).

\bibitem{Lozovik76}
Yu.E. Lozovik, V.I. Yudson,
Sov. Phys. JETP {\bf 44}, 389
(1976).

\bibitem{Fukuzawa90}
T. Fukuzawa, S. Kano, T. Gustafson, T. Ogawa,
Surf. Sci. {\bf 228}, 482
(1990).

\bibitem{Butov01}
L.V. Butov, A.L. Ivanov, A. Imamoglu, P.B. Littlewood, A.A. Shashkin, V.T. Dolgopolov, K.L. Campman, A.C. Gossard,
Phys. Rev. Lett. {\bf 86}, 5608
(2001.)

\bibitem{Butov95}
L.V. Butov, A.Zrenner, G. Abstreiter, A.V. Petinova, K. Eberl,
Phys. Rev. B {\bf 52}, 12153
(1995).

\bibitem{Dzyubenko96}
A.B. Dzyubenko, A.L. Yablonski,
Phys. Rev. B {\bf 53}, 16355
(1996).

\bibitem{Butov99}
L.V. Butov, A.A. Shashkin, V.T. Dolgopolov, K.L. Campman, A.C. Gossard,
Phys. Rev. B {\bf 60}, 8753 (1999).

\bibitem{Butov01a}
L.V. Butov, A. Imamoglu, K.L. Campmana, A.C. Gossard,
J. Exp. Theor. Phys. {\bf 92}, 260
(2001).

\bibitem{Kowalik-Seidl11}
K. Kowalik-Seidl, X.P. V{\"o}gele, F. Seilmeier, D. Schuh, W. Wegscheider, A.W. Holleitner, J.P. Kotthaus,
Phys. Rev. B {\bf 83}, 081307(R) (2011).

\bibitem{Schinner13}
G.J. Schinner, J. Repp, K. Kowalik-Seidl, E. Schubert, M.P. Stallhofer, A.K. Rai, D. Reuter, A.D. Wieck, A.O. Govorov, A.W. Holleitner, J.P. Kotthaus,
Phys. Rev. B {\bf 87}, 041303(R) (2013).

\bibitem{Lozovik97}
Yu.E. Lozovik, A.M. Ruvinskii,
Sov. Phys. JETP {\bf 85}, 979 (1997).

\bibitem{Butov01b}
L.V. Butov, C.W. Lai, D.S. Chemla, Yu.E. Lozovik, K.L. Campman, A.C. Gossard,
Phys. Rev. Lett. {\bf 87}, 216804 (2001).

\bibitem{Lozovik02}
Yu.E. Lozovik, I.V. Ovchinnikov, S.Yu. Volkov, L.V. Butov, D.S. Chemla,
Phys. Rev. B {\bf 65}, 235304 (2002).

\bibitem{Wilkes16}
J Wilkes, E.A. Muljarov,
New J. Phys. {\bf 18}, 023032 (2016).

\bibitem{Gorbunov13}
A.V. Gorbunov, V.B. Timofeev,
Solid State Commun. {\bf 157}, 6
(2013).

\bibitem{High13}
A.A. High, A.T. Hammack, J.R. Leonard, Sen Yang, L.V. Butov, T. Ostatnick{\'y}, M. Vladimirova, A.V. Kavokin, T.C.H. Liew, K.L. Campman, A.C. Gossard,
Phys. Rev. Lett. {\bf 110}, 246403 (2013).

\bibitem{Elliott}
R.J. Elliott, R. Loudon,
J. Phys. Chem. Solids {\bf 8}, 382
(1959);
{\bf 15}, 196
(1960).

\bibitem{Hasegawa}
H. Hasegawa, R.E. Howard,
J. Phys. Chem. Solids {\bf 21}, 179
(1961).

\bibitem{Gor'kov68}
L.P. Gor'kov, I.E. Dzyaloshinskii,
Sov. Phys. JETP {\bf 26}, 449 (1968).

\bibitem{Lerner80}
I.V. Lerner, Yu. E. Lozovik,
Sov. Phys. JETP {\bf 51}, 588 (1980).

\bibitem{Kallin84}
C. Kallin, B.I. Halperin, Phys. Rev. B {\bf 30}, 5655 (1984).

\bibitem{Feldman87}
J. Feldmann, G. Peter, E.O. G{\"o}bel, P. Dawson, K. Moore, C. Foxon, R.J. Elliott, Phys. Rev. Lett. {\bf 59}, 2337 (1987).

\bibitem{Hanamura88}
E. Hanamura, Phys. Rev. B {\bf 38}, 1228 (1988).

\bibitem{Andreani91}
L.C. Andreani, F. Tassone, F. Bassani, Solid State Commun. {\bf 77}, 641 (1991).

\bibitem{Maialle93}
M.Z. Maialle, E.A. de Andrada e Silva, L.J. Sham, Phys. Rev. B {\bf 47}, 15 776 (1993).

\bibitem{suppl_mat}
Supplementary materials present IMX spatial profiles and spectra at different $B$ and $P_{\rm ex}$, external ring role, bulk emission subtraction, and theory details.

\bibitem{Ivanov06}
A.L. Ivanov, L.E. Smallwood, A.T. Hammack, Sen Yang, L.V. Butov, A.C. Gossard, Europhys. Lett. 73, 920 (2006).

\bibitem{Hammack09}
A.T. Hammack, L.V. Butov, J. Wilkes, L. Mouchliadis, E.A. Muljarov, A.L. Ivanov, A.C. Gossard,
Phys. Rev. B {\bf 80}, 155331 (2009).

\bibitem{Kuznetsova12}
Y.Y. Kuznetsova, J.R. Leonard, L.V. Butov, J. Wilkes, E.A. Muljarov, K.L. Campman, A.C. Gossard,
Phys. Rev. B {\bf 85}, 165452 (2012).

\bibitem{Butov04}
L.V. Butov, L.S. Levitov, A.V. Mintsev, B.D. Simons, A.C. Gossard, D.S. Chemla, Phys. Rev. Lett. {\bf 92}, 117404 (2004).

\bibitem{Rapaport04}
R. Rapaport, G. Chen, D. Snoke, S.H. Simon, L. Pfeiffer, K. West, Y. Liu, S. Denev, Phys. Rev. Lett. {\bf 92}, 117405 (2004).

\bibitem{Chen05}
G. Chen, R. Rapaport, S.H. Simon, L. Pfeiffer, K. West, Phys. Rev. B {\bf 71}, 041301(R) (2005).

\bibitem{Haque06}
M. Haque, Phys. Rev. E {\bf 73}, 066207 (2006).

\bibitem{Yang10}
Sen Yang, L.V. Butov, L.S. Levitov, B.D. Simons, A.C. Gossard, Phys. Rev. B {\bf 81}, 115320 (2010).

\bibitem{Sivalertporn12}
K. Sivalertporn, L. Mouchliadis, A.L. Ivanov, R. Philp, E.A. Muljarov, Phys. Rev. B {\bf 85}, 045207 (2012).


\end{references}
\end{document}